\documentclass[preprint,showpacs,preprintnumbers,amsmath,amssymb]{revtex4}

\usepackage{graphicx}
\usepackage{dcolumn}
\usepackage{bm}
\usepackage{color}

\newcommand{\para}{\|}

\begin{document}
\preprint{APS/123-QED}

\title{Angular Momentum Transport and Particle Acceleration \\ during Magnetorotational Instability in a Kinetic Accretion Disk}

\author{Masahiro Hoshino}
 \email{hoshino@eps.s.u-tokyo.ac.jp}
 \affiliation{Department of Earth and Planetary Science, University of Tokyo, Tokyo 113-0033, Japan}

\date{\today}

\begin{abstract}
Angular momentum transport and particle acceleration during the magnetorotational instability 
(MRI) in a collisionless accretion disk are investigated using three-dimensional 
particle-in-cell (PIC) simulation.  We show that the kinetic MRI can provide not only 
high energy particle acceleration but also enhancement of angular momentum transport.
We find that the plasma pressure anisotropy inside the channel flow with 
$p_{\para} > p_{\perp}$ induced by active magnetic reconnection suppresses the onset of 
subsequent reconnection, which in turn leads to high magnetic field saturation  
and enhancement of Maxwell stress tensor of angular momentum transport.   
Meanwhile, during the quiescent stage of reconnection the plasma isotropization
progresses in the channel flow, and the anisotropic plasma with $p_{\perp} > p_{\para}$ 
due to the dynamo action of MRI outside the channel flow contributes to rapid 
reconnection and strong particle acceleration. 
This efficient particle acceleration and enhanced  
angular momentum transport in a collisionless accretion disk may explain 
the origin of high energy particles observed around massive black holes.
\end{abstract}

\pacs{52.35.Vd, 52.65.Rr, 95.30.Qd, 97.10.Gz}
\maketitle

A fundamental obstacle in our understanding of the universe is the need to
explain the angular momentum redistribution in an accretion disk 
gravitationally rotating around a central object.  
It has been proposed that magnetorotational instability (MRI) is the most 
efficient mechanism for transporting angular momentum outward with 
inward mass motion \citep{BalHaw91,BalHaw98}.
Based on magnetohydrodynamic (MHD) simulations 
\citep[e.g.,][]{Hawley91,Hawley92,Matsumoto95,Stone96,Sano04},
it has been asserted that a weakly magnetized disk with an outwardly 
decreasing angular velocity gradient can provide angular momentum transport 
at a greatly enhanced rate by generating MHD turbulence.

While the MHD framework is successful in explaining the ``collisional'' 
accretion disks, it is also important to study the dynamnics of ``collisionless''
accretion disks for some classes of astrophysical objects \citep{Quat02}.
The accretion disk around the super-massive black hole Sagittarius A* at the 
center of our Galaxy is believed to be in a collisionless plasma state.  
This is because the accretion proceeds through a hot and low-density plasma 
in which the proton temperature is higher than the electron temperature 
\citep[e.g.,][]{Nara98}.
In addition to the non-equilibrium temperature between protons and electrons, 
nonthermal high-energy particles are observed 
\citep[e.g.,][]{Yuan03,Aharonian08,Chernyakova11,Kusunose12}.

Motivated by the observation of this collisionless accretion disk, 
\citet{Sharma03,Sharma06} studied the MRI, including the effect of pressure 
anisotropy.  Since the MRI involves the process of magnetic field amplification/dynamo, 
the perpendicular pressure is expected to be enhanced in the double adiabatic 
approximation \citep{CGL56}, which in turn modifies the MHD wave behavior.  
Meanwhile, because of the MHD waves generated by pressure anisotropy instabilities 
\citep[e.g.,][]{Gary97}, pressure isotropization occurs during MRI evolution.
\citet{Riquelme12,Hoshino13} performed two-dimensional PIC simulations and confirmed 
the excitation of the mirror mode and the relaxation of pressure anisotropy studied by 
the previous fluid-based model \citep{Sharma06}.  In addition to the pressure 
anisotropy effect, the formation of a power-law energy spectrum during magnetic 
reconnection was pointed out.  

Although the previous two-dimensional PIC simulation showed the importance of 
the kinetic accretion disk, the plasma transport process is generally different 
depending on whether it is considered in two-dimensional or three-dimensional space. 
The most important determinant of 
the efficiency of angular momentum transport in the collisionless MRI remains an 
open question.  In this letter, we investigate for the first time the collisionless 
MRI using a three-dimensional PIC simulation, and argue that the angular momentum 
transport can be enhanced by the pressure anisotropy.

\begin{figure*}
\includegraphics[scale=0.45]{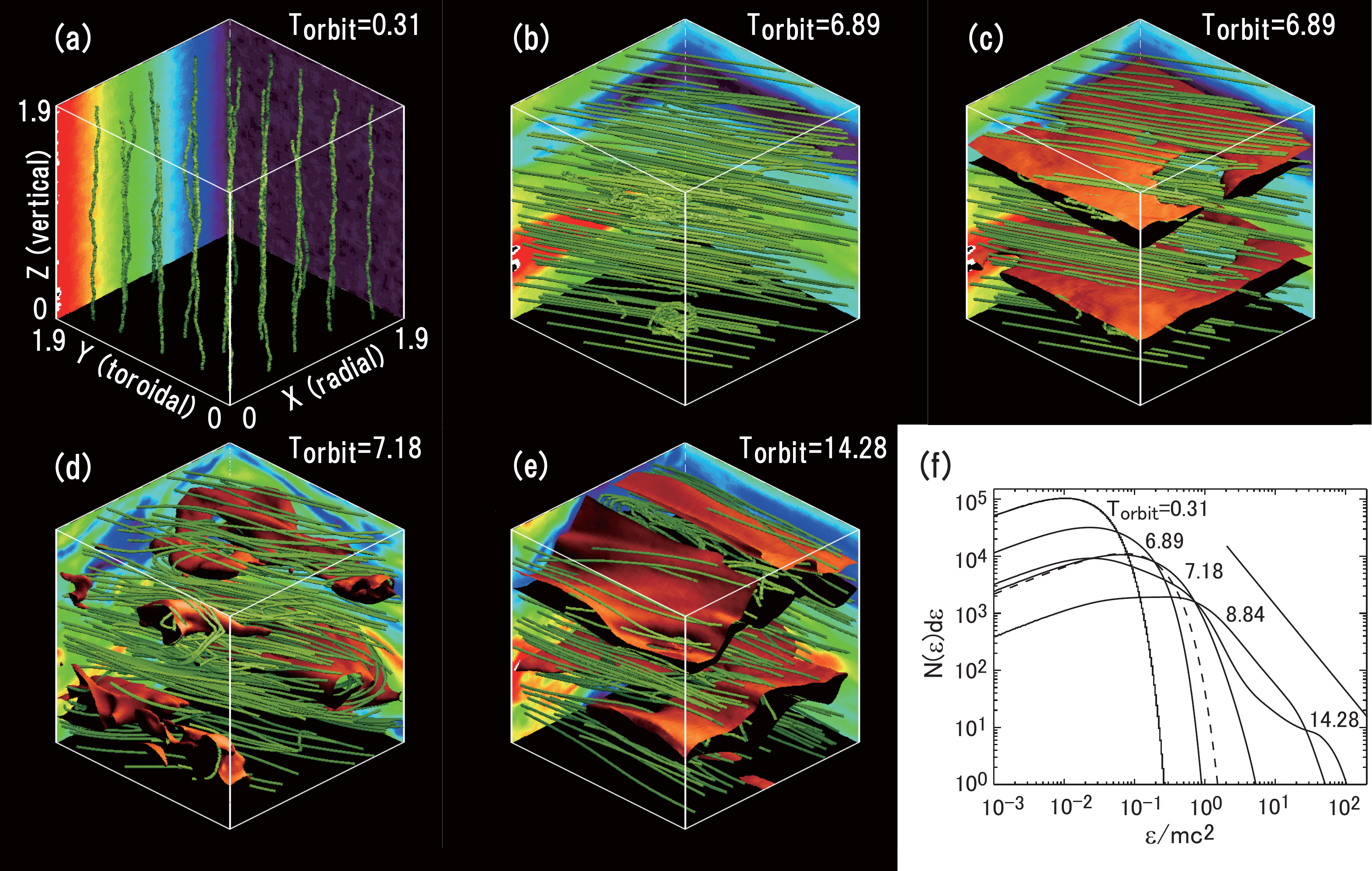}
\caption{Time evolution of the magnetorotational instability. Panels (a--b) show the 
magnetic field lines (greenish lines) and angular velocities in the background at 
$Y=y/\lambda=1.91$ and $X=x/\lambda=1.91$ (color contour), and panels (c--e) depict 
the high density regions as reddish curved planes.  The panels (b) and (c) are at
the same time stage. Panel (f): The energy spectra during the MRI at $T_{\rm orbit}=0.31$, 
$6.89$, $7.18$, $8.84$ and $14.28$.  The dashed line is a Maxwellian fitting for 
$T_{\rm orbit}=7.18$.}
\end{figure*}

To study the kinetic accretion disk in three-dimensional space, we performed a 
PIC simulation in a local frame rotating with angular velocity 
$\Omega_0 \vec{e}_z$ at a distance $r_0$ from the central object, and we include 
Coriolis, centrifugal, and gravitational forces in the equations of motion.  We 
used tidal expansion of the effective potential with a constant 
$q=-\partial \rm{ln} \Omega/\partial \rm{ln} r$ at $r_0$, where $q$ is  
$3/2$ for a Keplerian disk.  The equation of motion becomes
\begin{equation}
  \frac{d\vec{p}}{dt} = e(\vec{E} + \frac{\vec{v}}{c} \times \vec{B})
  -m \gamma(2 \vec{\Omega}_0 \times \vec{v}- 2q \Omega_0^2 x \vec{e}_x). 
\end{equation}
Our scheme was the same as that used in our previous two-dimensional MRI study, 
and we assumed that the local rotating velocity $\Omega_0 r_0$ is 
much smaller than the speed of light \citep{Hoshino13}.
We adopted the shearing box boundary condition established 
by MHD simulations \citep{Hawley95}.

For the initial condition, a drifting Maxwellian velocity distribution function was 
assumed in the local rotating frame with angular velocity $\Omega_0(r_0)$. 
The drift velocity in the $y$-direction $v_y(x)$ was given by 
$v_y(x)=r \Omega(r)-r \Omega_0(r_0) \simeq -q \Omega_0(r_0)x$, and the radial 
velocity $v_x$ and the vertical velocity $v_z$ were both zero.
In order to save CPU time, we set up the pair plasma, but the linear behavior of 
the MRI in the pair plasma was the same as that of ion-electron plasmas 
\citep{Hoshino13}.  A non-relativistic, isotropic plasma 
pressure with a high plasma $\beta = 8 \pi (p_+ + p_-)/B_0^2 = 1536$ was assumed, 
where the electron and positron gas pressures were related to the thermal velocities 
$v_{t\pm}$ by $p_\pm = (3/2) m_\pm n v_{t\pm}^2$.  The initial magnetic field was 
oriented purely vertical to the accretion disk; i.e., $\vec{B}=(0,0,B_0)$. 
The ratio of the cyclotron frequency to the disk angular velocity was fixed at 
$\Omega_{c\pm}/\Omega_0= \pm 10$, where $\Omega_{c\pm}=e_\pm B_0/m_\pm c$.
The grid size $\Delta$ was set to $ 2^{3/2}(v_{t\pm}/\Omega_{p\pm})$, 
where $\Omega_{p\pm}=\sqrt{8 \pi n e^2/m_\pm}$ is the pair plasma frequency.
The Alfv\'en velocity is defined as $V_A=B/\sqrt{8 \pi m_{\pm} n}$, so that
the plasma $\beta$ is equal to $3 v_{t\pm}^2/V_A^2$.  The parameters used were
$(V_A/\Omega_0)/\Delta=25$, $(v_{t\pm}/\Omega_{c\pm})/\Delta=56.4$, 
$V_A/c=6.25 \times 10^{-3}$.  $N_x$, $N_y$, and $N_z$ are the grid sizes in the
$x$, $y$, and $z$ directions, respectively, and we assumed $N_x = N_z = N_z = 300$ 
in this paper. 
$L_x=L_y=L_z=(N_x \Delta)/\lambda=1.91$ is the physical size normalized by 
$\lambda=2 \pi V_A/ \Omega_0$. 
The number of particles per cell was set to $N_p$/cell $=40$.  

Figure 1 shows the time evolution of the magnetic field lines (greenish lines) 
and the structure of the high density regions (sandwiched by the reddish curved 
planes).  Color contours in the background at $Y=1.91$ and $X=1.91$ show the 
angular velocity $v_y$ in the local rotating frame.  In the early stage at 
$T_{\rm orbit}=\Omega_0 t/2 \pi = 0.31$ in panel (a), the magnetic field lines are 
parallel to the $z$ axis, and the Keplerian motion/differential motion of $v_y$ can 
be seen as the color contour at $Y=1.91$, where the reddish (bluish) region
corresponds to a positive (negative) toroidal velocity.  As time passes, the 
vertical magnetic fields start to get distorted due to the MRI, 
and they are stretched out in the toroidal direction because of the 
Keplerian motion at $T_{\rm orbit}=6.89$ in panel (b).  This stretching motion 
can amplify the magnetic field and form two inward- and outward-flowing streams 
with a high plasma density and strong electric current, called the channel flow.  
The reddish regions sandwiched by two surfaces in panel (c) show the high density 
channel flow with $\rho \ge  \langle \rho \rangle + 2 \sigma_{\rho}$ 
where $\langle \rho \rangle$ and $\sigma_{\rho}$ are the average density and 
standard deviation of density distribution in the simulation domain, respectively. 

The amplification of the magnetic field stretched by the Keplerian motion may be 
balanced by the magnetic field dissipation caused by magnetic reconnection.  Panel 
(d) at $T_{\rm orbit}=7.18$ is the stage just after the onset of magnetic reconnection, 
and the break of the laminar channel flow seen at $T_{\rm orbit}=6.89$ can be observed.
After the first onset of reconnection, subsequent reconnection occurs intermittently 
in several different sites in the turbulent channel flows, and the formation
of the channel flow with a strong magnetic field by MRI dynamo and 
destruction by reconnection occurs repeatedly.

Panel (f) shows the time evolution of the energy spectra, where the horizontal 
and vertical axes are the particle energy normalized by the rest mass energy and  
number density $N(\varepsilon)$.  Before the first onset of magnetic reconnection 
at $T_{\rm orbit} = 6.89$, the plasmas are gradually heated from the initial cold 
Maxwellian plasma.  After the onset of reconnection at $T_{\rm orbit}=7.18$, we can 
clearly observe nonthermal particles above $\varepsilon/mc^2 > 0.2$.  
The dashed line, for reference, is a Maxwellian spectrum fitted 
by $T/mc^2=0.121$.  The nonthermal population continues to grow, 
and the spectrum can be approximated by a single power law function with 
$N(\varepsilon) \propto \varepsilon^{-3/2}$ at $T_{\rm orbit}=8.84$.
After $T_{\rm orbit}=9 \sim 10$, the spectrum slop becomes harder in
the high energy range from $\varepsilon/mc^2 \sim 10$ to $10^2$.
The spectrum hardening might be due to the stochastic, multiple 
reconnection process \citep{Hoshino12}, but note that 
the maximum attainable energy in the system, whose gyro-radius is almost 
the same as the simulation box size, is $\varepsilon/mc^2 \sim 10^2$.
Then the spectrum deformation might be related to the accumulation of high 
energy particles around the maximum attainable energy.
As already discussed by the previous 2D PIC simulations 
\citep{Riquelme12,Hoshino13}, the pressure anisotropy with $p_{\perp}>p_{\para}$
is generated in our 3D simulation by the MRI dynamo (see Figure 3), 
which can contribute to rapid reconnection \citep{Chen84} 
and particle acceleration \citep{Hoshino13}.

Let us take a look at the history of kinetic and magnetic field energies in the
top panel of Figure 2.  The energies are normalized by the initial magnetic field 
energy.  As time goes on, both the kinetic and magnetic field energies increase, 
but the rapid increase of the magnetic field energy can be observed at around 
$T_{\rm orbit} \sim 6$, and the instantaneous plasma $\beta$ becomes of the order of 
unity.  Our PIC simulation in the local rotating system has been carried out using 
the open shearing box boundary condition \citep{Hawley95}, and the plasmas can acquire 
their energies by accretion toward the center of gravity.  Around $T_{\rm orbit} \sim 8$, 
the total magnetic field energy reaches to its maximum, and then starts to decrease 
until $T_{\rm orbit} \sim 9$.  After $T_{\rm orbit} \sim 9$, both the magnetic field 
and kinetic energies remain almost constant with fluctuations.

\begin{figure}
\includegraphics[scale=0.5]{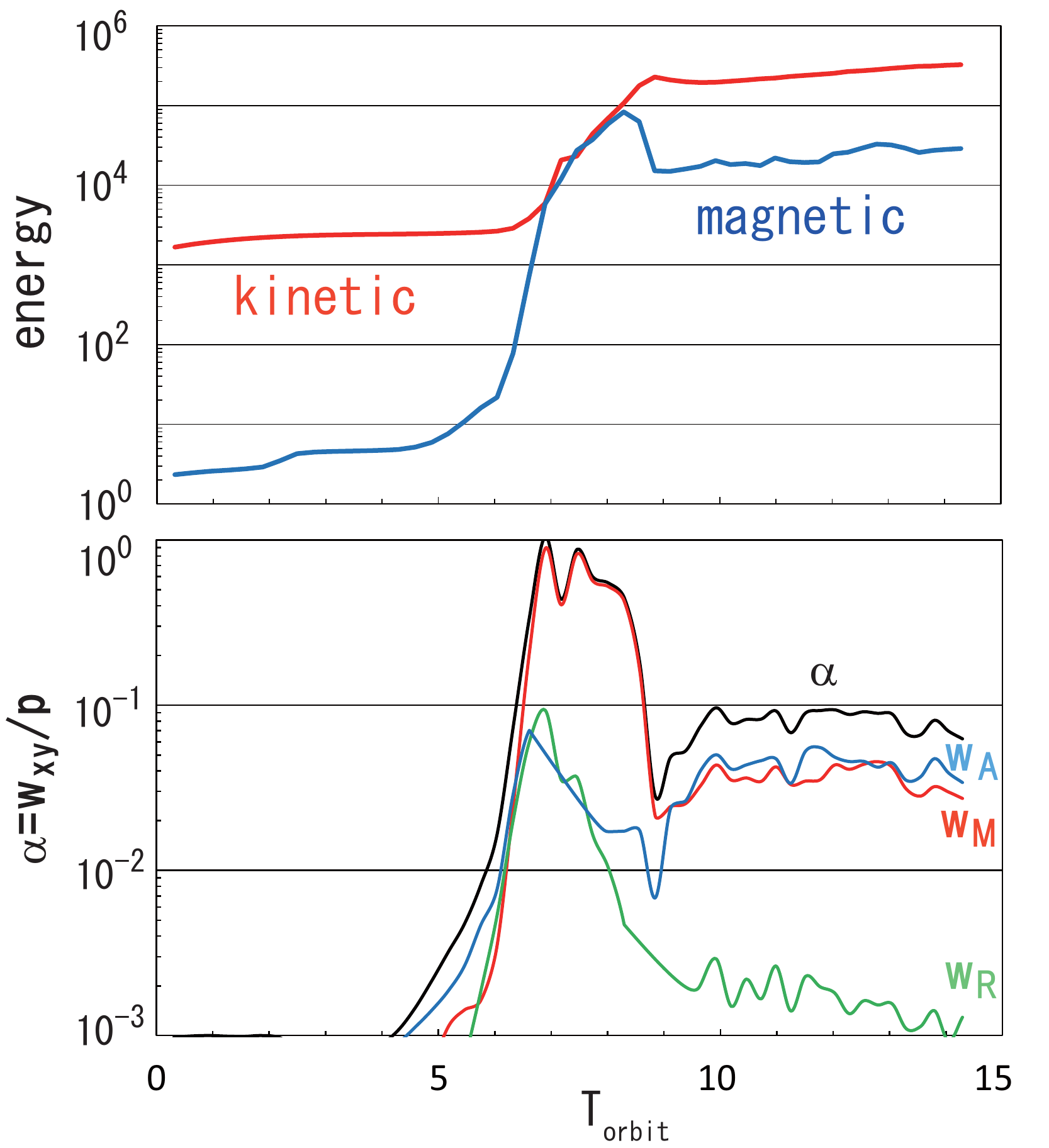}
\caption{(Top) Time evolution of the kinetic and magnetic field energies.  
(Bottom) The $\alpha$ parameter (black) and the contribution of Reynolds stress 
$w_R/p$ (green), Maxwell stress $w_M/p$ (red), and anisotropy stress $w_A/p$ (blue).}
\end{figure}

This time evolution is basically similar to our previous 2D PIC simulation result 
\citep{Hoshino13}.  In the early nonlinear stage, we observe the formation of 
channel flows with the amplification of the magnetic field and the subsequent 
break of the channel flow by reconnection.  The main difference, however, is seen 
in the late nonlinear stage with the turbulent/intermittent reconnection after 
$T_{\rm orbit} \ge 9$.  In our previous 2D simulation, after the first onset of 
reconnection, the channel flows are destroyed, and a couple of large magnetic 
islands are formed in the system.  In this 3D simulation, however, the channel 
flows are preserved beyond the first onset of reconnection, and magnetic 
reconnection occurs in several different locations with the dynamic motion.  
The turbulent/intermittent reconnection in the late phase can be seen in 3D MHD 
simulations as well \citep{Hawley91,Hawley92,Matsumoto95,Stone96,Sano04}. 

The most intriguing result in our kinetic MRI simulation is the enhancement of the 
angular momentum transport.  The bottom panel in Figure 2 shows the time evolution 
of parameter $\alpha$, which is used in the standard accretion disk model 
\citep{Shakura73} and can be defined as $\alpha = w_{xy}/p$, where $p$ and $w_{xy}$ 
are the volume-averaged, instantaneous plasma pressure, and stress tensor, 
respectively.  The stress tensor $w_{xy}$, which is related to the energy 
dissipation rate in the system, can be calculated as follows:
\begin{equation}
  w_{xy}=\rho v_x (v_y + q \Omega_0 x)-\frac{B_x B_y}{4 \pi} + 
         \frac{(p_{\para}-p_{\perp})}{B^2}B_x B_y.
\end{equation}
The terms on the right-hand side represent the Reynolds ($w_{R}$), Maxwell ($w_{M}$), 
and anisotropy ($w_{A}$) stresses, respectively \citep{Sharma06}.  
During the active reconnection phase between $7 < T_{\rm orbit} < 9$,
we found that $\alpha$ reached $O(1)$ with $w_{M} > w_{R} \sim w_{A}$, 
and during the late stage of $T_{\rm orbit}>9$, $\alpha \sim O(10^{-1})$ with 
$w_{M} \sim w_{A} > w_{R}$, which suggests a much more efficient angular momentum 
transport than the one discussed previously with $\alpha \sim O(10^{-3}-10^{-2})$
\citep{Hawley95,Sano04}.  Note that $\alpha$ in some simulation results were 
normalized by the initial pressure $p_0$ instead of the instantaneous pressure $p$.

Since the parameter $\alpha$ is approximated by,
\begin{equation}
  \alpha =\frac{w_{xy}}{p} \sim \left(-\frac{2 B_x B_y}{B^2} \right)
              \left(\frac{B^2/8 \pi}{p} \right) \sim \frac{1}{\beta},
\end{equation}
the enhancement of parameter $\alpha$ is related to a higher saturation of the 
magnetic field, which can be determined from the balance between the magnetic field
amplification due to the MRI dynamo and magnetic field dissipation by reconnection.  
The dissipated magnetic field energy is deposited as thermal energy.
If the onset of reconnection requires a high magnetic field in the collisionless
system, then the plasma $\beta$ becomes small and a large $\alpha$ can be realized.

\begin{figure}
\includegraphics[scale=0.45]{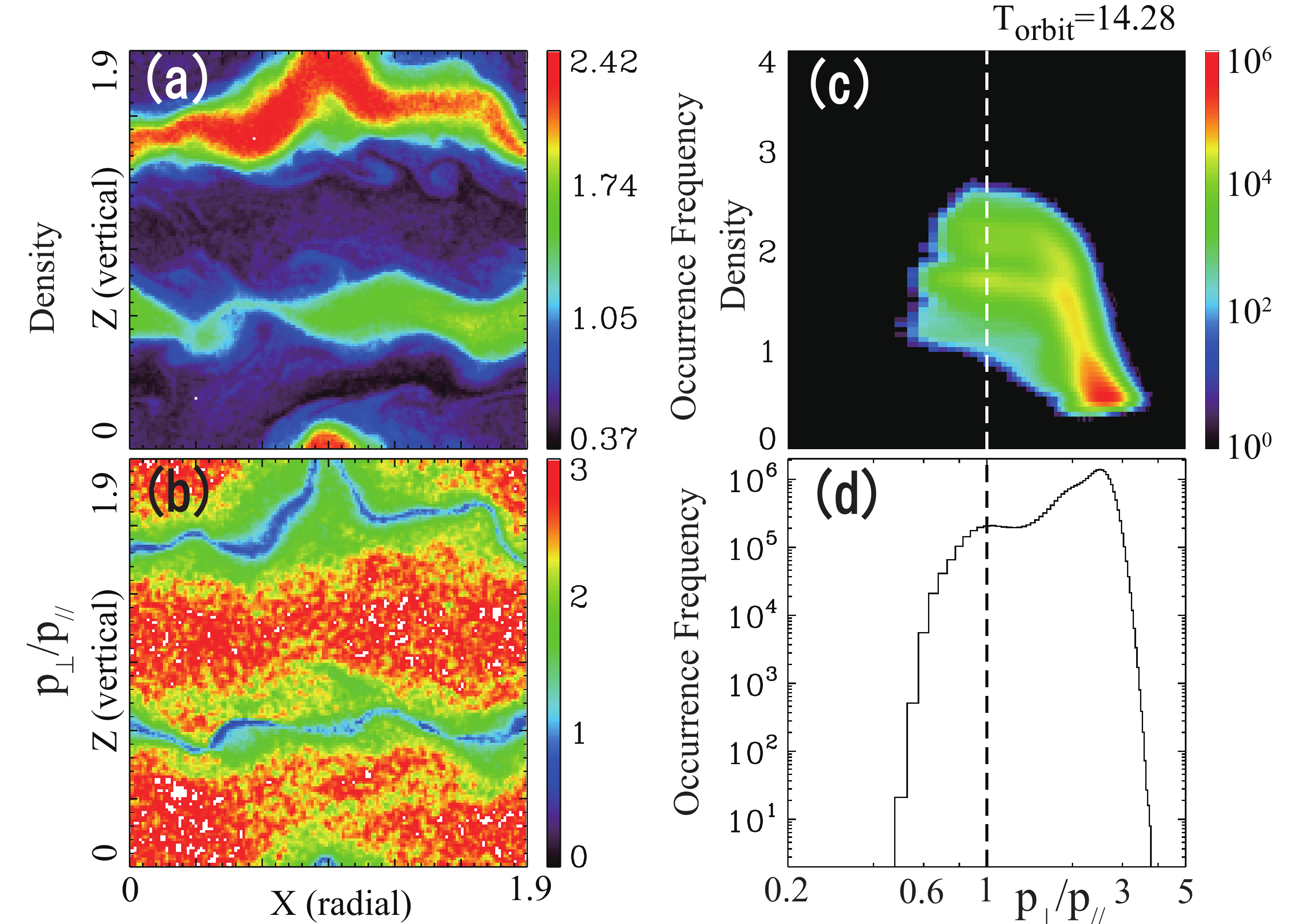}
\caption{(a) A slice of the three-dimensional density in $x-z$ plane at $Y=0.96$, 
and (b) a slice of the pressure anisotropy $p_{\perp}/p_{\para}$ in the same plane.  
The color scales are shown on the right.  (c) The color contour of the occurrence 
frequency in $(\rho, p_{\perp}/p_{\para})$ plane, and (d) the histogram of the 
occurrence frequency as a function of $p_{\perp}/p_{\para}$ in the logarithmic
scale.}
\end{figure}

To understand the dynamics of reconnection, we focus on the pressure anisotropy.  
Panels (a) and (b) in Figure 3 at $T_{\rm orbit}=14.28$ show, respectively, two 
dimensional $(x,z)$ slices of plasma density and pressure anisotropy 
$p_{\perp}/p_{\para}$ at the position $Y=0.96$. The higher density regions 
around $Z=0.64$ and $1.60$ correspond to the so-called channel flows where the 
magnetic field polarity changes.  At the center of the channel flow, one can see
$p_{\perp}/p_{\para} \le 1$, while $p_{\perp}/p_{\para} > 1$ for the other regions.
Panel (c) shows the relationship between the plasma density and $p_{\perp}/p_{\para}$,
and panel (d) is the histogram of occurrence frequency.  
This behavior can be basically understood using the double adiabatic equation of 
state with $p_{\perp}/\rho B = const.$ and 
$p_{\para}B^2/\rho^3 =const.$
\citep{Sharma06,Riquelme12,Hoshino13}.  The production of $p_{\perp}/p_{\para} > 1$
is simply due to the magnetic field amplification of the MRI dynamo, and the 
formation of $p_{\para}/p_{\perp} > 1$ 
is because of magnetic reconnection at the 
center of the channel flow, where the total B is dissipated while the plasma density 
is compressed.
In the kinetic perspective, it is known that the pressure anisotropy can be produced 
by the Alfv\'enic beams along the plasma sheet boundary, which are emanating from 
the magnetic diffusion region \cite{Hoshino98}.

The onset of magnetic reconnection is still a controversial issue, but the linear 
growth rate of the collisionless tearing mode under the pressure anisotropy would be 
sufficient for our argument.  
This is given, for example by simplifying Eq.(40) in \citep{Chen84}, 
as follows:
\begin{equation}
  \frac{{\rm Im}(\omega)}{k v_{th}} \simeq 
     \left( \frac{p_{\perp}}{p_{\para}}-1 \right) +
     \left( \frac{r_g}{\delta} \right)^{3/2} 
     \left( \frac{1 - k^2 \delta^2}{k \delta} \right),
\end{equation}
where $k$, $\delta$, $r_g$, and $v_{th}$ are the wave number, the thickness of the 
current sheet, gyro-radius, and thermal velocity, respectively.  At the 
saturation stage, $v_{th} \le c$, $B/B_0 \sim 230$, and $\delta/\Delta \ge 10$.  We then 
obtain the estimates of $(r_g/\delta)^{3/2} \le 0.089$ and 
$(1-k^2 \delta^2)/k \delta \sim O(1)$.  On the other hand,
the pressure anisotropy is $1-p_\perp/p_\para < 0.5$ from Figure 3 (d).  Therefore, 
it is highly possible that the successive reconnection in the channel flow is 
suppressed by anisotropic plasma of $p_{\perp}/p_{\para} < 1$, which is formed by the 
preceding reconnection.  The kinetic magnetic reconnection involves a deterrent effect 
to the successive magnetic dissipation, and as a result, the high magnetic field is 
realized before the onset of reconnection.

\begin{figure}
\includegraphics[scale=0.4]{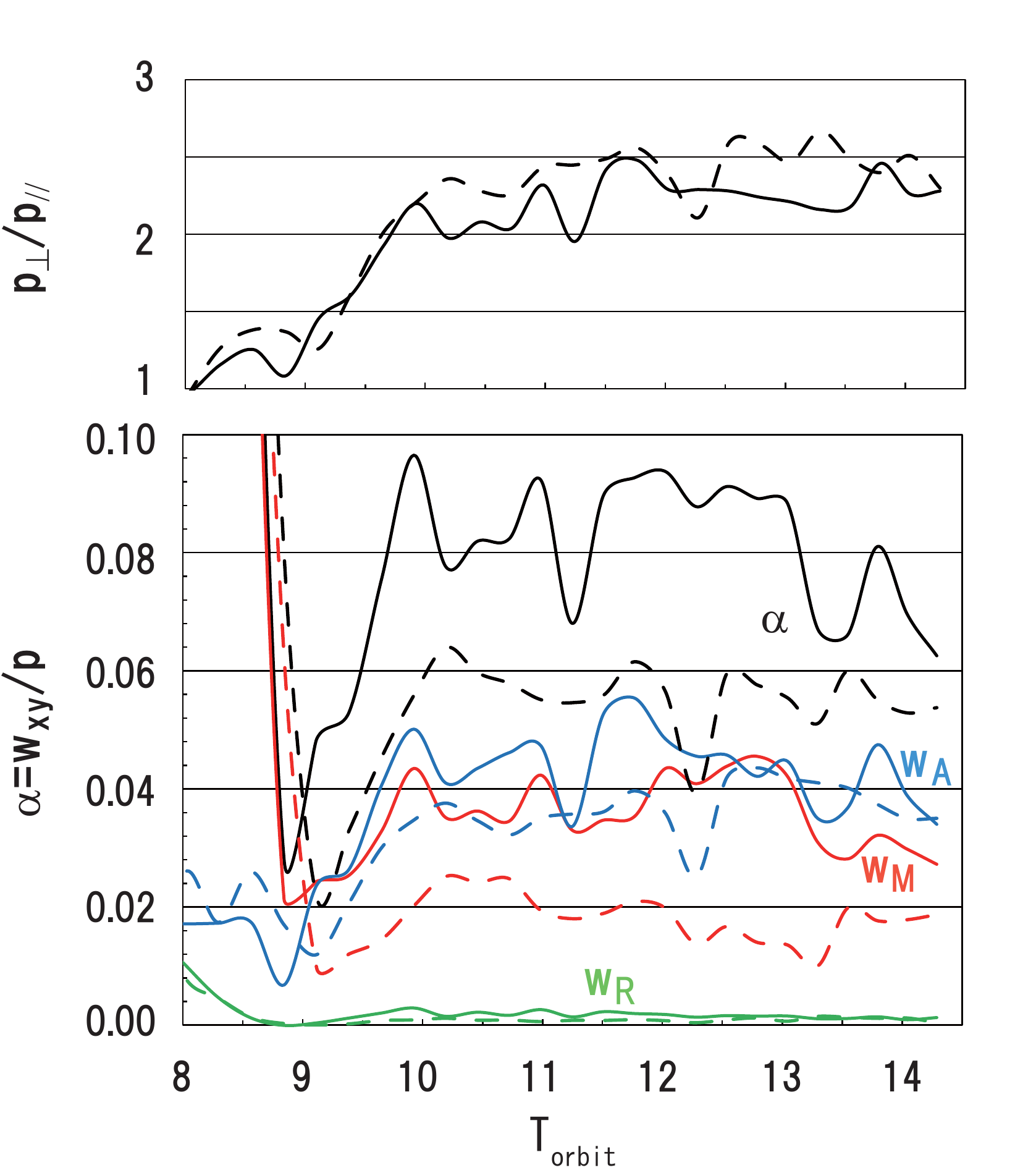}
\caption{(Top) Time evolution of the volume integrated pressure anisotropy.
The dashed and solid lines show the cases with and without the isotropization model, 
respectively.
(Bottom) The $\alpha$ parameter (black) and the contribution of Reynolds stress $w_R/p$
(green), Maxwell stress $w_M/p$ (red), and anisotropy stress $w_A/p$ (blue). }
\end{figure}

To confirm the enhancement of $\alpha$ parameter by the $p_{\para} > p_{\perp}$ effect
in the channel flow, 
we performed another simulation including an isotropization model for an anisotropic 
plasma after the formation of the channel flow, $T_{\rm orbit} > 6.89$.  
In this model, we added a weak external random magnetic field $\delta \vec{B}$ 
in Eq.(1) only for the channel flow region which is roughly characterized by
the weak magnetic field region of $|B/B_0| < 50$.  Namely, we used the equation 
of motion given by,
\begin{equation}
\frac{d\vec{p}}{dt} = e \left(
    \vec{E} + \frac{\vec{v}}{c} \times (\vec{B}+ \delta \vec{B}) \right)
    +(\rm{other~forces}),
\end{equation}
where we assumed the white noise $\delta B$ with $|\delta B|/|B|=2.5$.  

Shown in the top panel of Figure 4 is the time history of the volume integrated pressure 
anisotropy $p_{\perp}/p_{\para}$. The dashed and solid lines correspond to the cases 
with and without the above isotropization model, respectively.  
As we expected, one can find that the $p_{\perp}/p_{\para}$ with the isotropization
model in the channel flow is larger than that without the isotropization, because
the anisotropic plasma with $p_{\para} > p_{\perp}$ in the channel flow can be
reduced by the isotropization model.  

Under this weak isotropization in the channel flow, let us study the time history
of the $\alpha$ parameter and the Reynolds $w_R/p$, Maxwell $w_M/p$, and anisotropy 
stress $w_A/p$  with and without the isotropization model in the bottom panel.  
The dashed and solid lines show the cases with and without the isotropization 
model, respectively. One can find that $\alpha$ parameter can be reduced under the 
isotropization model.  The magnitude of the anisotropy stress $w_A/p$ does not 
change between two cases with and without the isotropization model, while the change 
of the Maxwell stress $w_M/p$ becomes large under the isotropization model.  
This suggests that the isotropization in the channel 
flow plays an important role on the magnetic field generation during the MRI dynamo. 

In summary, we have investigated for the first time a three-dimensional, collisionless 
MRI in a local rotating system, and have shown that an anisotropic pressure of 
$p_{\para}/p_{\perp} > 1$ is maintained in the channel flow during the MRI, which leads 
to high magnetic field saturation and an enhanced $\alpha$ parameter.  During the 
quiescent stage of reconnection the isotropization of the anisotropic plasma progresses 
in the channel flow.  After the plasma isotropization the anisotropic plasma with 
$p_{\perp}/p_{\para} > 1$ outside the channel flow region may contribute to a rapid 
reconnection and nonthermal particle generation \citep{Hoshino13}.

\bibliography{mri3d}

\end{document}